\documentclass[12pt,preprint]{aastex}
\usepackage{amstext,amsmath,natbib}

\shorttitle{Testing Cluster Mass Indicators}
\shortauthors{JUETT ET AL.}

\begin{document}

\title{Testing the Reliability of Cluster Mass Indicators with a
Systematics Limited Dataset}

\author{Adrienne~M.~Juett\altaffilmark{1},
David~S.~Davis\altaffilmark{2,3}, Richard Mushotzky\altaffilmark{4}}

\altaffiltext{1}{NASA Postdoctoral Fellow, Laboratory for X-Ray
Astrophysics, Code 662, NASA/Goddard Space Flight Center, Greenbelt,
MD 20771; ajuett@milkyway.gsfc.nasa.gov}
\altaffiltext{2}{Department of Physics, University of Maryland,
Baltimore County, 1000 Hilltop Circle, Baltimore, MD 21250, USA}
\altaffiltext{3}{CRESST and the Astroparticle Physics Laboratory,
NASA/GSFC, Greenbelt, MD 20771, USA david.s.davis@nasa.gov}
\altaffiltext{4}{Laboratory for X-ray Astrophysics, Code 662,
NASA/Goddard Space Flight Center, Greenbelt, MD 20771, USA
richard@milkyway.gsfc.nasa.gov}

\begin{abstract}
We present the mass--X-ray observable scaling relationships for
clusters of galaxies using the {\em XMM-Newton} cluster catalog of
\citeauthor{smk+08}. Our results are roughly consistent with previous
observational and theoretical work, with one major exception.  We find
2--3 times the scatter around the best fit mass scaling relationships
as expected from cluster simulations or seen in other observational
studies.  We suggest that this is a consequence of using hydrostatic
mass, as opposed to virial mass, and is due to the explicit dependence
of the hydrostatic mass on the gradients of the temperature and gas
density profiles.  We find a larger range of slope in the cluster
temperature profiles at $r_{500}$ than previous observational studies.
Additionally, we find only a weak dependence of the gas mass fraction
on cluster mass, consistent with a constant.  Our average gas mass
fraction results argue for a closer study of the systematic errors due
to instrumental calibration and analysis method variations.  We
suggest that a more careful study of the differences between various
observational results and with cluster simulations is needed to
understand sources of bias and scatter in cosmological studies of
galaxy clusters.
\end{abstract}


\section{Introduction}
Studies of clusters of galaxies provide for a variety of cosmological
tests \citep[see][for a recent review]{v05}.  The precision of these
tests is limited by the accuracy and precision of the scaling
relations used to transform X-ray observables, such as temperature or
luminosity, into cluster mass measurements
\citep[e.g.,][]{ram08,vbe+09}.  Improved modeling of the complex
physics at work in clusters has led to better study of the sources of
bias and scatter in the mass scaling relationships.  These models
suggest that the hydrostatic equilibrium assumption used to calculate
total cluster masses from X-ray data can lead to underestimates on the
order of 10--20\% \citep[e.g.,][]{nvk07}.  The discrepancy is
explained by the presence of nonthermal pressure support in clusters.
Theoretical studies have found that the expected scatter in mass
scaling relationships is dependent on the dynamical state of the
clusters under study \citep[e.g.,][]{kvn06}.  Relaxed systems should
show lower scatter around the best-fit relation than disturbed
clusters.

This information can be used to tailor observational programs,
particularly by limiting the bias and scatter in the mass scaling
relationships.  One suggestion is to use only relaxed clusters in
cosmological studies.  However for high redshift samples, relaxed
clusters make up only a minority of the cluster population \citep[see
  e.g.,][]{vbe+09}.  Alternatively, the right choice of mass proxy
could provide low scatter data.  \citet{kvn06} proposed a new X-ray
proxy for cluster mass, the parameter $Y_X$ which is the product of
X-ray temperature and gas mass.  The $Y_X$ parameter is related to the
total thermal energy of the cluster gas and is an X-ray analog to the
Sunyaev-Zel'dovich (SZ) flux.  From cluster simulations, \citet{kvn06}
found that the scatter in the cluster mass--$Y_X$ scaling law is not
only lower than for other commonly used mass proxies, but also shows
little dependence on cluster dynamical state.

Observational studies of the $Y_X$ parameter as a mass proxy have been
limited in the number of clusters used and the statistical quality of
the data \citep{app07,vbe+09}.  The statistical error was on the order
of the measured scatter, making it difficult to determine the
intrinsic scatter in the relationship.  Additionally, previous
observational work has focused solely on a limited number of relaxed
systems.  A larger and more sensitive observational study of clusters
will better test the $Y_X$ mass proxy and the results of cluster
simulations.  In this Letter, we use the data from a recent high
signal-to-noise {\em XMM-Newton} survey of nearby clusters to test the
usefulness of various mass proxies.  We assume
$H_0=70$~km~s$^{-1}$~Mpc$^{-1}$, $\Omega_M=0.3$, and
$\Omega_\Lambda=0.7$.

\section{Data Analysis}
Our cluster sample was taken from \citet{smk+08}, which presented the
projected temperature, abundance, and surface brightness profiles for
70 clusters found by fitting {\em XMM-Newton} data (see that work for
details of the spectral analysis).  While not an unbiased sample, the
large number of clusters does provide a wide range of cluster
properties to study.  It is not limited to relaxed and/or hot ($>
5$~keV) clusters.  
The selection criteria used by \citet{smk+08} excluded highly
asymmetric clusters and those with strong substructure (e.g., A115,
A754).  However, some merging clusters were included in the sample
(e.g., A520, the Bullet Cluster).

We used the projected profiles to determine the gas density and
three-dimensional temperature profiles following the procedure of
\citet{vkf+06}.  The same gas density and temperature models were used
with some simplification.  The lower spatial resolution of our data
did not constrain the second $\beta$-model component required to fit
the {\em Chandra} data in the cluster center, therefore we did not
include it.  We also compared fits with and without steepening at
large radii and excluded the steepening when it did not produce a
significantly better fit of the data.  For the temperature fits, we
took into account projection effects and used the spectral temperature
weighting formalism of \citet{v06}.  When cooling in the core was not
obvious we fixed the cool component parameters to produce a flat
temperature profile in the center.

We calculated the total cluster mass distribution for each cluster
assuming hydrostatic equilibrium and the radii ($r_{2500}$ and
$r_{500}$) where the cluster density equaled 2500 and 500 times the
critical density at the cluster redshift. We do not include clusters
where the calculated $r_{2500}$ and $r_{500}$ values extend beyond the
radial coverage of our data.  We find that 60/70 clusters have data
extending to at least $r_{2500}$ and 28/70 have data out to $r_{500}$.
We determined the gas mass ($M_{g}$) and total cluster mass ($M$)
enclosed by $r_{2500}$ and $r_{500}$ for each cluster.  We also
calculated the average spectral temperature ($T_X$) within the 0.15--1
$r_{500}$ radial range using the formulation of \citet{v06}.

Uncertainty intervals were obtained from Monte Carlo simulations.  We
simulated surface brightness and projected temperature distributions
by scattering the observed data according to the measurement
uncertainties found in \citet{smk+08}. The simulated data were fit with
 gas density and temperature models and a full analysis performed
to determine $M$, $M_{g}$, and $T_X$.  The uncertainties were
obtained from their distribution in the simulated data.  For values
evaluated at $r_{500}$ ($r_{2500}$), the uncertainty includes the
uncertainty on $r_{500}$ ($r_{2500}$).

\section{Comparison of Scaling Relations}

We determined the best-fit scaling relationships at $r_{500}$ using
$E(z)^{n} M = C(X/X_0)^{\alpha}$ for $X = T_{X}$, $M_{g}$, and
$Y_{X}$.  We fixed $n=$ 1, 0 and 2/5, as consistent with our
cosmology, and $X_0=$ 5~keV, $4\times10^{13} \, M_{\odot}$ and
$3\times10^{14} \, M_{\odot} \, {\rm keV}$ for the $T_{X}$, $M_{g}$
and $Y_{X}$ fits, respectively.  We also present the best-fit
relationship between the gas mass fraction, $f_{g} = M_{g}/M$, and
$M$, as characterized by the equation $f_{g} = f_{g,0} + \alpha
\log_{10}[M/10^{15}M_{\odot}]$ \citep[see][]{vbe+09}.

We use the BCES fitting routines\footnote{Routines available at
http://www.astro.wisc.edu/$\sim$mab/archive/stats/stats.html} which
provide a linear regression algorithm that allows for intrinsic
scatter and nonuniform measurement errors in both variables
\citep{ab96}.  We find that the Y$|$X and orthogonal slope estimators
provide the most significant (and consistent) results compared to
other methods, including bisector.  We therefore only present the
results of the Y$|$X and orthogonal methods for each of our fits (see
Table~\ref{tab:fit}).  We also include the best-fit results when the
relationship slope, $\alpha$, is fixed at the expected value from
self-similarity.

For the $M-X$ relationships we estimate the intrinsic scatter using a
generalized form of the estimated scatter, $\delta M/M$, used by
\citet{vkf+06}:
\begin{equation}
\left( \frac{\delta M}{M} \right)^2 = \frac{1}{N-2} \sum \frac{[M_i -
C(X_i/X_0)^{\alpha}]^2 - \Delta M_i^2}{M_i^2} ,
\end{equation}
where $\Delta M_i$ are the measurement errors.  Similarly for the
$f_{g}-M$ relationship, we calculate the scatter $\delta f_{g}/f_{g}$
by:
\begin{equation}
\left( \frac{\delta f_g}{f_g} \right)^2 = \frac{1}{N-2} \sum
\frac{[f_{g,i} - (f_{g,0} + \alpha \log_{10}[M_i/10^{15}M_{\odot}])]^2 -
\Delta f_{g,i}^2}{f_{g,i}^2} ,
\end{equation}
where $\Delta f_{g,i}$ are the measurement errors.  To compare our
scatter with logarithmic scatter estimates
\citep[e.g.,][]{jhb+08,pca+09}, multiply the logarithmic estimates by
$\ln 10 = 2.30$.  Given the high statistical quality of our data, the
scatter 
is dominated by the intrinsic scatter.

\subsection{Mass$-$Temperature}
\begin{figure}
\epsscale{0.8}
\plotone{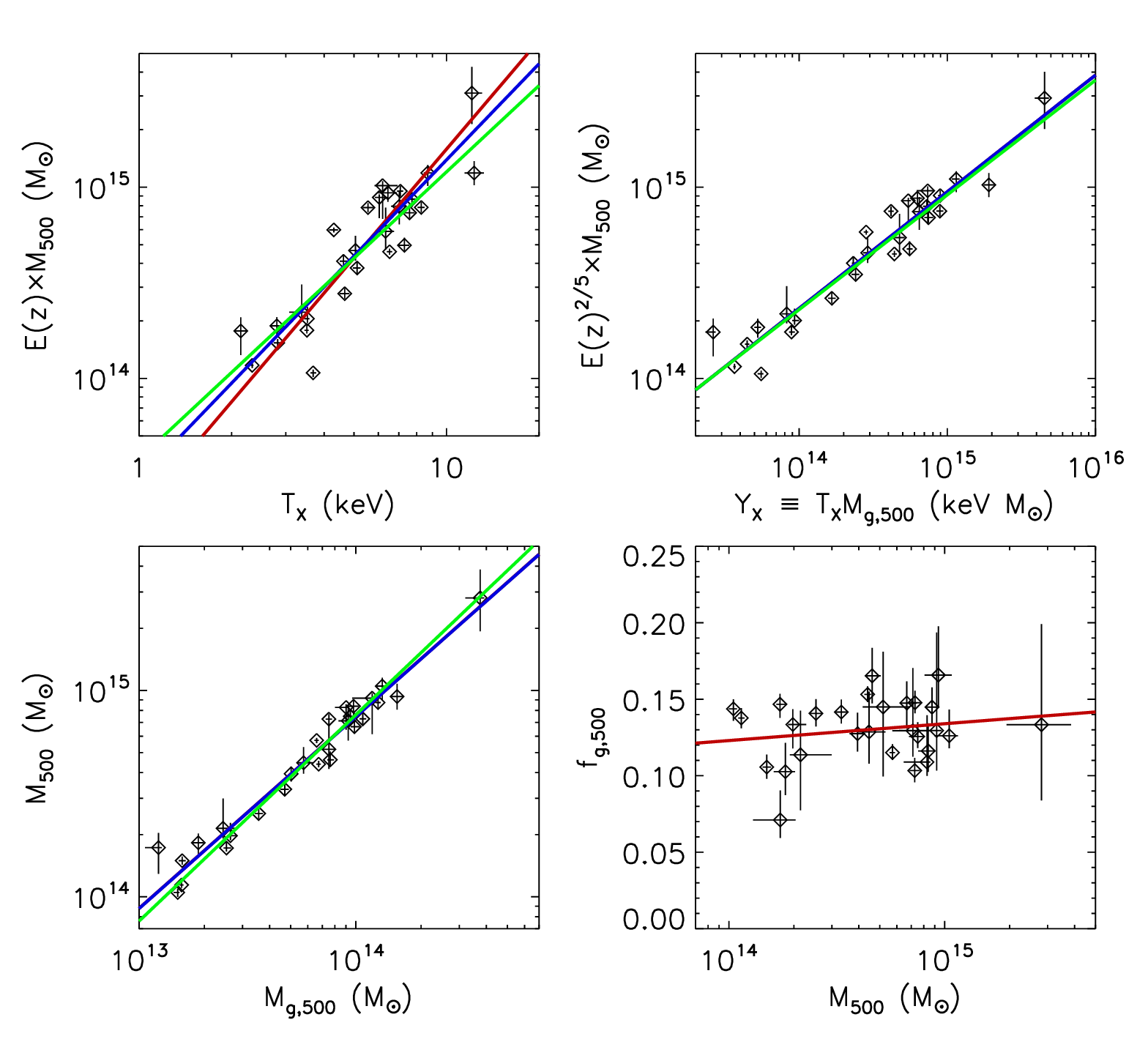}
\caption{{\em Upper Left:} Plot of X-ray spectral temperature, $T_X$,
and total cluster mass, $M$.  Overplotted are the best-fit power-law
relations using the BCES orthogonal slope estimator (red), the BCES
Y$|$X slope estimator (blue), and a fixed slope of $\alpha=1.5$
(green).  
{\em Upper Right:} Plot of $Y_X$ and $M$ with best-fit power-law
relations overplotted.  Color coding is the same as for the $M-T_{X}$
plot with $\alpha=0.6$.
{\em Lower Left:} Plot of cluster gas mass, $M_g$, and $M$ and
best-fit power-law relations overplotted.  Color coding is the same as
for the $M-T_{X}$ plot with $\alpha=1.0$.
{\em Lower Right:} Plot of gas mass fraction, $f_g$, and
$M$. Overplotted is the best-fit relation.}
\label{fig:fits}
\end{figure}

Table~\ref{tab:fit} and Figure~\ref{fig:fits} show the best-fit
parameters for the $M-T_{X}$ scaling relation.  Our best fit
($\log_{10} C =$ 14.64, $\alpha = 1.67$) is consistent with the
scaling relations found in other datasets \citep[$\log_{10} C =$
14.580 and 14.635, and $\alpha =$ 1.71 and 1.53,
respectively]{app07,vbe+09}.  \citet{app07} use a different definition
of $T_{X}$ which integrates the observed temperature profile over
0.15--0.75 $r_{500}$.  Since the average cluster temperature profile
falls at large radii \citep[see e.g.,][A. M. Juett et al. 2009, in
prep]{lm08}, this definition will produce higher values of $T_{X}$ and
subsequently lower values of the $M-T_{X}$ normalization.  We find
that $\langle T_{\rm X}(0.15-0.75 r_{500})/T_{\rm X}(0.15-1 r_{500})
\rangle = 1.06$, which translates into a reduction of $\log_{10} C$ of
0.04--0.05, enough to explain the discrepancy between our results and
\citet{app07}.

When compared to theoretical calculations
\citep{kvn06,nkv07,nvk07,jhb+08}, our results are consistent when the
hydrostatic mass is considered, but our normalization is lower by
$\approx20$\% when compared to scaling results that use the true
cluster mass (14.70--14.75).  This is a well known result and is
likely due to non-thermal pressure support that is not accounted for
in the hydrostatic mass estimate \citep[e.g.,][]{nkv07}.  The biggest
difference between our results and previous studies is the difference
in scatter around the best-fit scaling relation.  We find a scatter of
43\% which, due to our systematics limited dataset, can be attributed
to the intrinsic scatter of our sample.  Other studies, of both
observed and theoretical cluster samples, have found significantly
lower values for the intrinsic scatter, $\approx10-25$\%
\citep{vkf+06,app07,nkv07,nvk07,jhb+08}.

\subsection{Mass$-$Gas Mass}
The gas mass--total mass scaling relationship has been previously
characterized in two ways: a powerlaw scaling between $M$ and $M_{g}$,
and more recently by a linear scaling of $f_{g}$ and the logarithm of
$M$.

Our $M-M_{g}$ results ($\log_{10} C = 14.503\pm0.018$), are in
agreement with the normalization found by
\citet[14.542$\pm$0.015]{app07} considering the small formal
errors. But our best-fit slope is steeper (0.93$\pm$0.05 vs
0.80$\pm$0.04).  Again we find that theoretical models that use the
true cluster mass have higher predicted normalizations but our results
are consistent when hydrostatic mass estimates are used
\citep{kvn06,nkv07}.  The slope estimate is consistent with
theoretical results for both true and hydrostatic mass estimates.  Our
15\% scatter is close to the $\approx$10\% scatter found in both
observational and theoretical work.  Interestingly, our results match
the combination of normalization, slope, and scatter found in the
simulations of \citet{nkv07} when hydrostatic mass is used, however we
differ from their {\em Chandra} observational results.  Our slope is
larger (0.93 vs 0.81) while our normalization is lower (14.503 vs
14.59).  This discrepancy may be due to differences in the calibration
of the instruments that has been previously noted
\citep[e.g.,][]{smk+08}.

We find that the normalization of the $f_{g}-M$ relationship,
$0.134\pm0.005$, is consistent with the value of
\citet[0.130$\pm$0.007]{vbe+09}.  Our data are consistent with a
constant slope over the range of mass considered, while the
\citet{vbe+09} result prefers a reduction in $f_{g}$ for lower mass
clusters.  The \citet{vbe+09} result is consistent with gas mass
fraction results from groups of galaxies \citep{svd+09}.
\citet{app07} suggested that the gas mass fraction may be constant
above $2-3 \times 10^{14} M_{\odot}$, and then dropping at lower
masses.  This would explain both our result and the lower gas mass
fractions seen in groups of galaxies.

The difference in mean gas mass fraction between the \citet{vkf+06}
work, a subset of the \citet{vbe+09} sample, and ours is 10--20\%.  At
$r_{500}$, we find a mean $f_{g} = 0.1323\pm0.0019$, compared to
0.110$\pm$0.002 for the \citet{vkf+06} sample (adjusted to account for
differences in cosmology).  If we restrict our analysis to clusters
with $kT > 5$~keV, the difference is $\approx$10\%, 0.138$\pm$0.003
for our work compared to 0.123$\pm$0.003 \citep{vkf+06}.

Our results are also close to those found by \citet{ars+08}.  At
$r_{2500}$, \citet{ars+08} found a mean cluster gas mass fraction of
0.113$\pm$0.003 for clusters with $kT > 5$~keV and low redshifts ($z <
0.15$).  For their full sample ($kT > 5$~keV but all redshifts) they
find a mean $f_{g} = 0.1104\pm0.0016$.  \citet{vkf+06} noted that
their mean $f_{g}$ at $r_{2500}$ was significantly less
(0.091$\pm$0.002, a $\sim$25\% difference) than an earlier (but
consistent) Allen et al.~sample.  We find $f_{g,2500} =
0.1057\pm0.0005$ for clusters with $kT > 5$~keV, a 4\% difference with
the \citet{ars+08} results and a $\approx$15\% difference with the
\citet{vkf+06} work.

\subsection{Mass$-Y_X$}
\citet{kvn06} suggested that a lower scatter ($<10$\%) proxy for
cluster mass is the parameter $Y_{X} = T_{X} M_{g}$.
Table~\ref{tab:fit} gives our best-fits for the $M-Y_{X}$ scaling
relation.  Our best-fit normalization (14.657$\pm$0.018) is consistent
with the results of \citet[14.653$\pm$0.015 when renormalized]{app07}
and \citet[14.684$\pm$0.015]{vbe+09}.  When comparing with theoretical
results, we find consistency with hydrostatic mass results (14.645)
but not true cluster mass \citep[14.712; e.g.][]{nkv07}.  The slope of
the best-fit $M-Y_{X}$ compares well with both observational and
theoretical studies.

The largest difference comes in the measured scatter in the $M-Y_{X}$
relation.  We find a scatter of $\approx22$\%, much larger than the
$<10$\% level expected from simulations.  While other observational
studies have not found as large a scatter \citep[e.g.,][]{app07}, we
note that ours is the first study to be systematics limited and
includes twice the number of objects.  Thus we are able to measure the
scatter without a significant contribution from the statistical error.
In theoretical work, scatter increases to 8--20\% when hydrostatic
masses are used compared to true masses \citep{nkv07,jhb+08}.

\section{Correlation of Deviation with Cluster Properties}\label{sec:why}
Scatter in the mass--X-ray observable relationships is an important
contributor to the total error budget in cosmological studies of
clusters \cite[see e.g.][]{vbe+09}.  Given our large sample of
clusters, we can study what factors are most important in producing
scatter in these relationships which can then be used to refine
cosmological studies to reduce the scatter.

One possible cause of the observed scatter is the dynamical state of
the cluster.  Relaxed clusters are expected to better follow the
assumptions of hydrostatic equilibrium.  Disturbed clusters may have
additional pressure and energy inputs due to the merging events and
the additional complication of asymmetric geometries
\citep[e.g.,][]{nvk07}.

\begin{figure}
\epsscale{0.8}
\plotone{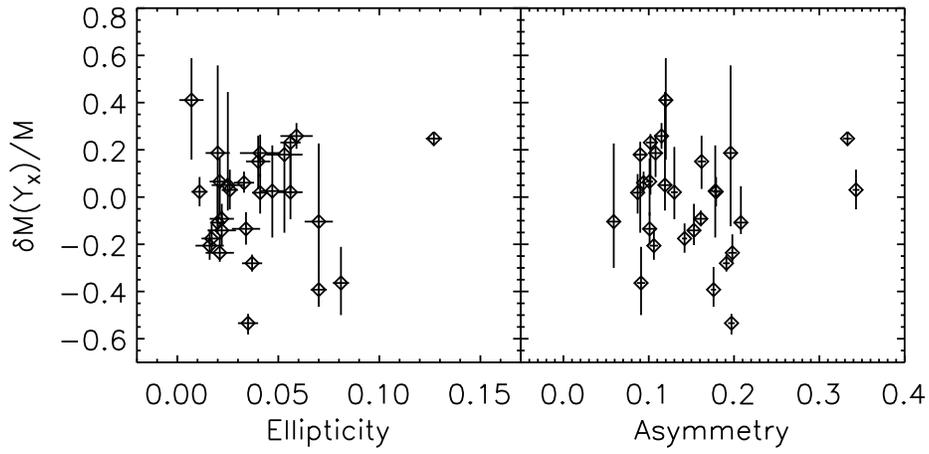}
\caption{Plot of deviation ($\delta M(Y_X)/M$) around the
best-fit $M-Y_{X}$ scaling relationship versus measures of the cluster
dynamical state, ellipticity ({\em left panel}) and asymmetry ({\em
right panel}).  No correlation is found.}
\label{fig:scat}
\end{figure}

We looked for correlation between the deviation of the calculated
hydrostatic mass, $M_i$, from the expected mass given the best-fit
scaling relationship, $M(X_i)$, with measures of the ellipticity and
asymmetry in the cluster images (see Figure~\ref{fig:scat}).  We
identify the deviation as $\delta M(X)/M = [M_i - M(X_i)]/M_i$.  The
ellipticity and asymmetry were calculated following the work of
\citet{hbh+07}.  Relaxed clusters should have low ellipticity and
asymmetry values, while disturbed systems will have higher ellipticity
and/or asymmetry values.  \citet{hbh+07} showed that ellipticity is
correlated with the P2/P0 power ratio and that asymmetry is related to
the P3/P0 power ratio \citep[see e.g.,][for a discussion of power
  ratios]{bt95,jhb+08}.  We find no correlation between the amount of
deviation and either ellipticity or asymmetry for any of our scaling
laws.

\begin{figure}
\epsscale{0.5}
\plotone{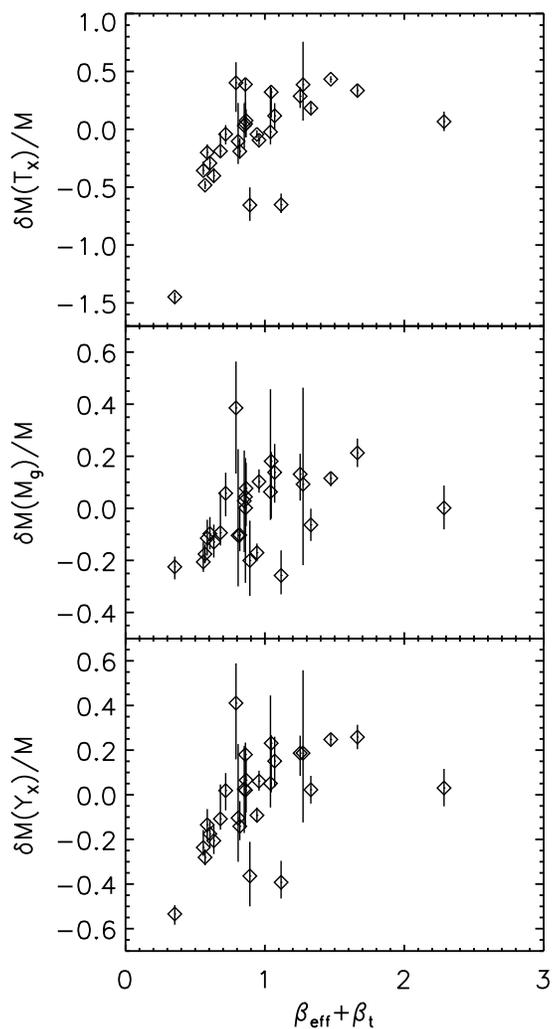}
\caption{Plot of deviation around the best-fit mass scaling relations,
$\delta M(X)/M$, versus $\beta_{eff} + \beta_t$.  The results for
the scaling relationships for $M-T_{X}$ ({\em top panel}), $M-M_{g}$
({\em middle panel}) and $M-Y_{X}$ ({\em bottom panel}) are given.
All show a correlation between deviation and $\beta_{eff} + \beta_t$.}
\label{fig:beta}
\end{figure}

We looked for other cluster properties that might influence the
deviation.  \citet{vkf+06} noted that implicit in the calculation of
hydrostatic mass, there is a dependence of the normalization of the
$M-T_{X}$ relation on the sum of the temperature gradient, $\beta_t =
(-1/3) \, d\log T/d\log r$, and gas density gradient, $\beta_{eff} =
(-1/3) \, d\log \rho/d\log r$ (see their Appendix A).  If the spread
in values is large enough, this dependence should cause a predictable
deviation around the best-fit relationship.  We find a strong
correlation between $\delta M(X)/M$ and $\beta_{eff} + \beta_t$ for
all scaling relationships (Figure~\ref{fig:beta}).  The Spearman rank
correlation coefficients were 0.62, 0.53, and 0.61 for $X=T_{X}$,
$M_{g}$, and $Y_{X}$, corresponding to probabilities of 0.9987,
0.9943, and 0.9986, respectively."

The \citet{vkf+06} cluster sample has a narrow distribution of
$\beta_{eff} + \beta_t$ values.  Our sample however, shows a large
variation in the temperature gradient.  For $\beta_t$, we find a mean
of 0.53 and a standard deviation of 0.50.  The range of temperature
gradient values is a reflection of the variation of temperature
profiles at large radii in our cluster sample \citep{smk+08}.  The gas
density gradient has a narrower distribution with a mean $\beta_{eff}
= 0.63$ and a standard deviation of 0.10.  Our gas density gradient
results are in good agreement with those found in the REXCESS study
\citep[$\beta_{ne}$ (0.3--0.8 $r_{500}$) = 0.60$\pm$0.10][]{cpb+08}.

\section{Discussion}
Our {\em XMM-Newton} survey of galaxy clusters \citep{smk+08},
provides a large sample and high signal-to-noise data to study the
scaling relationships between cluster mass and cluster X-ray
temperature, gas mass, gas mass fraction, and the mass proxy $Y_X$.
Our fits are in good agreement with other observational and
theoretical work, with one major caveat.  We find a significantly
larger scatter around the best-fit relationships than has been
previously seen.

The scatter around the best-fit scaling relationships is 2--3 times
that found in most other observational and theoretical work.
Interestingly, a recent study of the scaling relationship between the
Compton y-parameter from SZ studies and cluster masses obtained from
gravitational lensing also showed large scatter in the $M_{GL}-T_{X}$
and $M_{GL}-Y$ relationships \citep[41\% and 32\%,
  respectively;][]{msr+09}, however we note a difference in
methodology between their work and ours (\citet{msr+09} used a fixed
clustercentric radius of 350kpc, while we work in mass scaled units).
While other work has suggested that disturbed clusters will show more
scatter than relaxed clusters \citep[e.g.][]{kvn06}, those authors
used a visual classification of their simulated clusters, rather than
a quantitative determination, making comparison difficult.  Within our
sample, we find no correlation between measures of the cluster
dynamical state (ellipticity and asymmetry) and deviation around the
best-fit scaling relationships.

Only one cluster property, the combination $\beta_{eff} + \beta_{t}$,
showed a strong correlation with deviation.  Since hydrostatic mass
estimates, like those used here, depend on this explicitly, the result
should be expected \citep[e.g.][]{vkf+06}.  Our data are the first to
show this dependence due to the large sample size and range of cluster
properties included.  There is no reason to suspect that true cluster
masses will show such a dependence given the bias (and scatter) in
hydrostatic mass determinations found in simulations where the true
cluster masses are known \citep[see e.g,][]{nvk07}.

One question we must ask is how reliable are our determinations of the
temperature profiles, whose wide range dominates the measurement of
$\beta_{eff} + \beta_{t}$.  To check the reliability of the background
modeling and spectral fitting procedure, \citet{smk+08} compared the
profile of A1795 with published {\em Chandra}, {\em XMM-Newton}, and
{\em Suzaku} temperature profiles and found no significant difference.
They also found reproducibility for three clusters with multiple
observations.  An initial comparison of our average cluster
temperature profile is in good agreement with previous results both in
overall shape and expected scatter (see A. M. Juett et al.~2009, in prep).
The results of \citet{lm08} also suggest that cluster temperature
profiles, while generally showing a falling profile at large radii, do
show a range of profiles.  Followup observations are needed of the
most unusual systems to confirm our results.

Assuming our temperature profile range is indicative of the cluster
population, we then need to ask how does this result affect
cosmological studies using clusters.  First, a larger scatter should
be taken into consideration when discussing systematic errors in
cosmological studies \citep[see e.g.][]{vbe+09}.  However, given other
error sources, it is not clear that the mass scaling relationship
scatter would be the dominant error contributor.

It may be possible to correct for the scatter from theoretical studies
of cluster properties but it is unclear if present models are
consistent with our results.  \citet{nkv07} find little variation in
the temperature profile at $r_{500}$ in their simulations, although
these are limited to their relaxed subsample.  If a more thorough
study of the simulations is not able to reproduce the observed cluster
variation, that may point to some missing physics needed to better
describe the conditions within clusters.

Another result we would like to highlight is the comparison of our
average gas mass fraction with previous results.  The differences
between our work and others \citep[e.g.,][]{vkf+06,ars+08} range from
5--20\%.  This is comparable with expected systematic differences
between the instruments and analysis methods, but is significantly
larger than the statistical errors typically quoted.  In our opinion,
a study of the expected systematics due to (1) instrumental
calibration differences, and (2) data analysis methods must be
performed.  These issues are beyond the scope of this work, but will
be addressed in our future paper (A. M. Juett et al. 2009, in prep).

\begin{deluxetable}{lcccc}
\tablewidth{0pt} 
\setlength{\tabcolsep}{0.15in}
\tablecaption{Best-Fit Parameters of Powerlaw Fits}
\tablehead{\colhead{Relation\tablenotemark{a}} & 
\colhead{Fit Method\tablenotemark{b}} & \colhead{$\log_{10}C/f_{g,0}$} & 
\colhead{$\alpha$} & \colhead{Scatter} }
\startdata
M$-$T & Y$|$X  & 14.64$\pm$0.03 & 1.67$\pm$0.16 & 0.420 \\
      & Orth & 14.63$\pm$0.03 & 1.89$\pm$0.20 & 0.430 \\
      & Fix  & 14.63$\pm$0.03 & 1.5           & 0.412 \\

M$-$Y$_{X}$ & Y$|$X  & 14.657$\pm$0.018 & 0.61$\pm$0.04 & 0.218 \\
            & Orth & 14.657$\pm$0.018 & 0.61$\pm$0.04 & 0.218 \\
            & Fix  & 14.646$\pm$0.018 & 0.60          & 0.212 \\

M$-$M$_{g}$ & Y$|$X  & 14.503$\pm$0.018 & 0.93$\pm$0.05 & 0.144 \\
            & Orth & 14.504$\pm$0.017 & 0.93$\pm$0.05 & 0.145 \\
            & Fix  & 14.484$\pm$0.013 & 1.0         & 0.149 \\

$f_{g}-$M & Y$|$X  & 0.134$\pm$0.005 & 0.011$\pm$0.013 & 0.104 \\
          & Orth & 0.134$\pm$0.005 & 0.011$\pm$0.013 & 0.104 \\

\enddata 
\tablenotetext{a}{The relationships are of the form $E(z)^{n} M =
C(X/X_0)^{\alpha}$ for the $M-T$, $M-Y_{X}$, and $M-M_{g}$ fits with
$n=1$, 2/5, and 0, and $X_0=5 \, {\rm keV}$, $3\times10^{14} \,
M_{\odot} \, {\rm keV}$ and $4\times10^{13} \, M_{\odot}$ for the T,
$Y_{X}$, and $M_{g}$ fits, respectively.  For the $f_{g}-$M fit, the
relationship is $f_{g} = f_{g,0} + \alpha
\log_{10}(M/10^{15}M_{\odot})$.}
\tablenotetext{b}{We use the BCES fitting package and present results
from the Y$|$X and orthogonal fitting methods.  In addition, we fit
the data with a fixed slope, $\alpha$, given by the expected
self-similar relationships ($\alpha = 1.5$, 1.0, and 0.6 for the T,
$Y_{X}$, and $M_{g}$ fits, respectively).}
\label{tab:fit}
\end{deluxetable}

\end{document}